\documentclass[twocolumn]{article}
\usepackage{PRIMEarxiv}

\usepackage{cite}
\usepackage{amsmath,amssymb,amsfonts}
\usepackage{algorithmic}
\usepackage{graphicx}
\usepackage{textcomp}

\usepackage{bm}
\usepackage{upgreek}
\usepackage{pifont}
\usepackage{tabularx}
\usepackage{hyperref}

\usepackage{xcolor} 
\definecolor{subsectioncolor}{RGB}{0,0,128}

\usepackage{titlesec}
\titlespacing*{\section}{0pt}{0pt}{0pt}
\titlespacing*{\subsection}{0pt}{0pt}{0pt}
\titlespacing*{\subsubsection}{0pt}{0pt}{0pt}

\usepackage{placeins}

\begin{document}
\title{Mitigating the Impact of Electrode Shift on Classification Performance in Electromyography-Based Motion Prediction Using Sliding-Window Normalization}

\author{
Taichi Tanaka\\
Department of Science Technology of Innovation\\
Nagaoka University of Technology\\
Nagaoka 940-2188, Japan\\
\texttt{taichi\_tanaka@stn.nagaokaut.ac.jp}\\
\and
Isao Nambu\\
Department of Electrical, Electronics and Information Engineering\\
Nagaoka University of Technology\\
Nagaoka 940-2188, Japan\\
\texttt{e-mail: inambu@vos.nagaokaut.ac.jp}\\
\and
Yasuhiro Wada\\
Department of Electrical, Electronics and Information Engineering\\
Nagaoka University of Technology\\
Nagaoka 940-2188, Japan\\
\\
}


\twocolumn[
\maketitle

\begin{abstract}
Electromyography (EMG) signals are used in many applications, including prosthetic hands, assistive suits, and rehabilitation. Recent advances in motion estimation have improved performance, yet challenges remain in cross-subject generalization, electrode shift, and daily variations. When electrode shift occurs, both transfer learning and adversarial domain adaptation improve classification performance by reducing the performance gap to -1\% (eight-class scenario). However, additional data are needed for re-training in transfer learning or for training in adversarial domain adaptation. To address this issue, we investigated a sliding-window normalization (SWN) technique in a real-time prediction scenario. This method combines z-score normalization with a sliding-window approach to reduce the decline in classification performance caused by electrode shift. We validated the effectiveness of SWN using experimental data from a target trajectory tracking task involving the right arm. For three motions classification (rest, flexion, and extension of the elbow) obtained from EMG signals, our offline analysis showed that SWN reduced the differential classification accuracy to -1.0\%, representing a 6.6\% improvement compared to the case without normalization (-7.6\%). Furthermore, when SWN was combined with a strategy that uses a mixture of multiple electrode positions, classification accuracy improved by an additional 2.4\% over the baseline. These results suggest that SWN can effectively reduce the performance degradation caused by electrode shift, thereby enhancing the practicality of EMG-based motion estimation systems.
\end{abstract}
\vspace{2ex}
]

\section{Introduction}
In the field of electromyography (EMG) for robot control and human motion prediction, researchers have investigated prosthetic hand control \cite{prosthetic-hand1, prosthetic-hand3_dnn}, gesture prediction \cite{transformer1_gesture, generalization_vae_gesture}, assist suit control \cite{assisut-suit1}, and rehabilitation \cite{rehabilitation2}. One study used a simple threshold for control \cite{prosthetic-hand1}, whereas another employed machine learning for control \cite{prosthetic-hand3_dnn}. In machine learning approaches, time-domain features such as integrated EMG, waveform length, and root mean square \cite{time-features} are extracted, while time-frequency features such as the spectrogram of the short-time Fourier transform \cite{STFT} and wavelets \cite{wavelet_over90persent} are utilized. The advantage of machine learning is that it enables automatic model construction for predicting more than three motion classes from handcrafted EMG features. However, handcrafted features pose challenges in determining the optimal window length for feature extraction and in selecting suitable feature combinations. Consequently, since 2016 most researchers have employed deep neural networks (DNNs) for human motion prediction \cite{prosthetic-hand3_dnn, electrode-shift_tl3}. DNNs automatically extract features from EMG signals, requiring only the window length to be set. Recent DNN models include  convolutional neural network-long short-term memory (CNN-LSTM) \cite{cnn-lstm1} and transformers \cite{transformer1_gesture}. CNN-LSTM is a hybrid architecture that uses CNN layers to extract local spatiotemporal features from EMG signals and LSTM layers to capture the temporal evolution of these features. In contrast, the transformer is a deep learning architecture based on self-attention that enables parallel processing of sequential data and has achieved state-of-the-art performance in tasks such as machine translation and language modeling.

Despite the high performance achieved by current machine learning and DNN methods, these approaches often require additional training data and calibration procedures, which impose significant operational burdens in real-world applications. Recent studies have focused on practical issues—such as cross-subject generalization \cite{generalization_vae_gesture, generalization_vae2, generalization_ada}, electrode shift \cite{electrode-shift_tl3, electrode-shift_tl1_mix_ref1, electrode-shift_tl2, electrode-shift_gesture, recalibration_ada_electrode-shift_daily-effects} and daily effects \cite{recalibration_ada_electrode-shift_daily-effects, recalibration_daily-effects2, recalibration_daily-effects3}—rather than solely improving machine learning performance \cite{over90persent1, over90persent2}. These issues arise because EMG amplitude varies with differences in muscle mass, electrode placement, muscle fatigue, and skin impedance (e.g., skin thickness and temperature). To address these challenges, various methods have been proposed, including transfer learning \cite{electrode-shift_tl1_mix_ref1, tl}, adversarial domain adaptation \cite{ada, recalibration_ada_electrode-shift_daily-effects}, cross-domain autoencoders for subject or session adaptation \cite{generalization_vae_gesture, generalization_vae2}, and re-calibration \cite{recalibration_ada_electrode-shift_daily-effects, recalibration_daily-effects2, recalibration_daily-effects3}. Transfer learning adapts a model trained on one task to a related task, thereby reducing the need for large amounts of labeled data and accelerating training by leveraging pre-trained representations. Adversarial domain adaptation uses adversarial training to align feature distributions between source and target domains, making the model more robust to domain shifts. Cross-domain autoencoders learn unified latent representations from different domains, enabling effective knowledge transfer. Re-calibration dynamically adjusts internal feature representations and output probabilities by re-training the model with estimated values, thus compensating for changes in EMG amplitude caused by muscle fatigue and ensuring consistently high performance even after prolonged use.

In electrode shift scenarios, both transfer learning and adversarial domain adaptation have been shown to reduce performance discrepancies. For example, A. Ameri et al. \cite{electrode-shift_tl1_mix_ref1} reported a 7\% classification error (eight-class scenario) under shifted conditions compared to 6\% under no-shift conditions (a difference of -1\%) using transfer learning with a CNN. Further, C{\^o}t{\'e}-Allard et al. \cite{recalibration_ada_electrode-shift_daily-effects} reported a 75.50\% classification accuracy (eleven-class scenario) under shifted conditions compared to 93.58\% under no-shift conditions (a difference of -18.08\%) using adversarial domain adaptation with a Spectrogram ConvNet. However, a major drawback of these methods is that they require additional data for re-training (in transfer learning) or for training (in adversarial domain adaptation and cross-domain autoencoders), which hinders their practical deployment. To overcome this issue, an alternative approach that does not require extra data is needed.

To improve cross-subject generalization, we proposed sliding-window normalization (SWN) \cite{mypaper}, which combines sliding-window processing with z-score normalization. Although the desired improvement in cross-subject generalization was not fully achieved, SWN increased classification accuracy from 56.2\% to 77.7\% (a 21.5\% improvement) for the same-subject model and from 41.1\% to 63.1\% (a 21.6\% improvement) for models of other subjects in three-class motion classification. Crucially, SWN achieves these improvements without the need for any additional calibration or training data. SWN improves performance by aligning EMG amplitude values within sliding windows and reducing variability, thereby enhancing signal consistency. Since electrode shift alters EMG amplitude due to changes in electrode position \cite{emg-amplitude-change1, emg-amplitude-change2}, SWN may mitigate its effects. Unlike transfer learning, adversarial domain adaptation, or cross-domain autoencoder approaches, which require extra data, SWN maintains robust performance solely through its normalization process. This data efficiency enables SWN to operate effectively in real-world settings, addressing a major limitation of current methods. Furthermore, as a normalization technique, SWN can be easily integrated with DNN approaches. Together, these advantages underscore the practical contribution of SWN in enhancing EMG-based applications for real-world deployment.

In summary, previous studies \cite{electrode-shift_tl1_mix_ref1, recalibration_ada_electrode-shift_daily-effects} have relied on additional data to achieve performance improvements, limiting their practical use. In contrast, the primary advantage of SWN is that it achieves comparable or improved performance without requiring any additional data. The aim of this study is to investigate the effect of sliding-window normalization (SWN) on reducing EMG-based real-time classification performance discrepancies while eliminating the need for additional data. To this end, in Section \ref{subsec:method_comparison} we compared the classification accuracy under electrode shift using a CNN-LSTM model for three motion classes with SWN against approaches using transfer learning (TL), adversarial domain adaptation (ADA), and a mixture of multiple electrode positions' data (MIX). In Section \ref{subsec:DNN_comparison_with_swn} we evaluated the benefits of integrating SWN with these DNN strategies.

\section{Methods}

\subsection{Data Acquisition}

\subsubsection{Subjects}
The Ethics Board of Nagaoka University of Technology approved this study in accordance with the Declaration of Helsinki (Number 2023-03-03). Seventeen right-handed men aged 21 to 24 participated in the experiment after being fully informed and providing their consent.

\subsubsection{Experiment}
In the experiment, data were collected under conditions with shifted electrode positions to evaluate SWN. Subjects performed a target marker tracking task with their right arm under three electrode position settings (left, center, and right) (Fig. \ref{fig:experiment}). The task involved 5 types of right arm movements (described in appendix) that elicited various motor dynamics (fast, slow, transitional) for evaluating classification performance. The task was conducted as 60 s trials repeated 4 times per session (60 trials total in 3 sessions). During the experiment, we recorded the positions of the right wrist, elbow, and shoulder, as well as the EMG signals from the muscles in the right forearm and right upper arm.

The positions of the right wrist, elbow, and shoulder were captured using a sensorless motion capture system (described in the appendix, sampling rate: 20 Hz). The system comprised three cameras (Cyber-shot RX100\uppercase\expandafter{\romannumeral 7}, Sony, Tokyo, Japan), a video capture board (CAM LINK PRO, Elgato, Fremont, California, USA), and a GPU (NVIDIA GeForce RTX 3070 Ti, NVIDIA, Santa Clara, CA, USA). A black cloth was placed between the air thread and the subject's right forearm to enhance the detection accuracy of the wrist and elbow joint positions.

EMG signals were recorded from 12 sites with the Trigno Lab Avanti system (Delsys, Natick, MA, USA, sampling rate: 2000 Hz). The recording sites included the biceps brachii ($\times$4), brachialis ($\times$1), brachioradialis ($\times$1), anconeus ($\times$1), triceps brachii (lateral head) ($\times$2), triceps brachii (long head) ($\times$2), and extensor carpi radialis longus ($\times$1). To thoroughly investigate the impact of electrode displacement on EMG signal acquisition and classification performance, electrode positions were systematically varied across three sessions. In Session 1, the electrode was placed at the center with maximal EMG amplitude. In Session 2, the electrode was shifted 2 cm to the right of the center, perpendicular to the arm's longitudinal axis, and in Session 3, it was shifted 2 cm to the left of the center. These controlled shifts in electrode placement allowed for a detailed evaluation of how varying electrode positions affect classification performance.

\begin{figure}[tbp]
\centering
\includegraphics[]{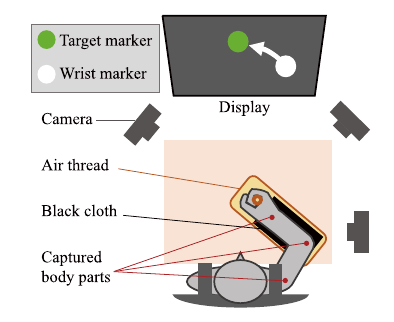}
\caption{An experimental environment. A black cloth was placed between the air thread and the subject's right forearm to enhance the detection accuracy of the wrist and elbow joint positions.}
\label{fig:experiment}
\end{figure}

\subsection{EMG Processing}
\begin{figure*}[htbp]
\centering
\includegraphics[width=\textwidth]{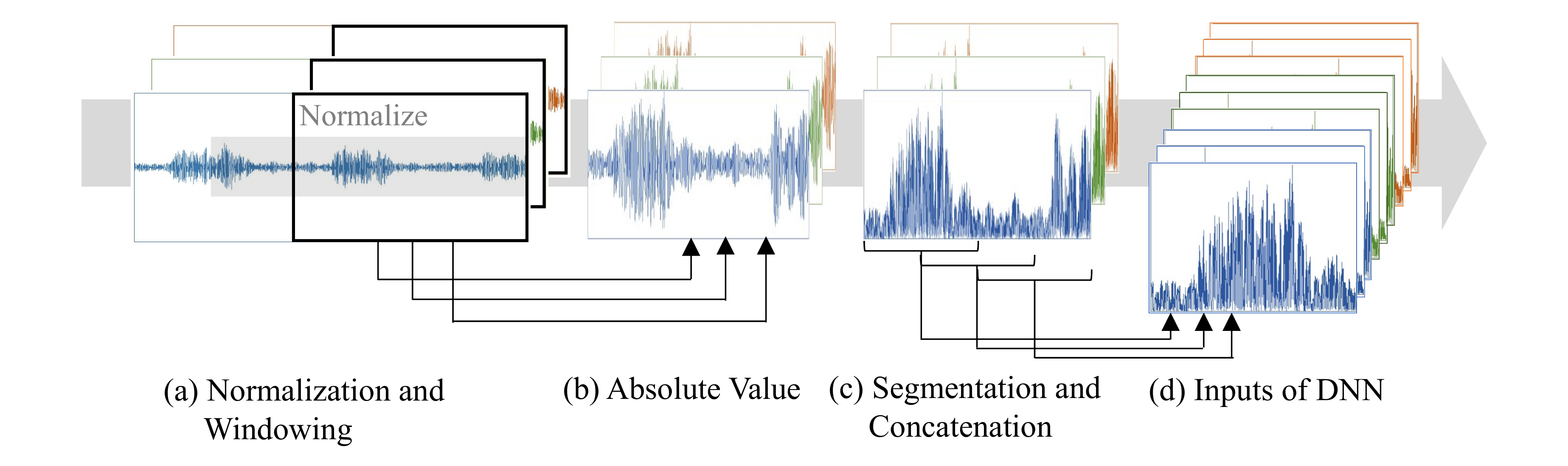}
\caption{The EMG Processing for real-time motion prediction.}
\label{fig:emg_processing}
\end{figure*}

For the DNN input, segmented absolute EMG signals were used to efficiently capture temporal variations. Initially, a 6th-order band-pass Butterworth filter (40-200 Hz) and decimation (2000 $\rightarrow$ 500 Hz) were applied first, followed by normalization over a fixed window length. Subsequently, signals of the desired feature extraction length were obtained, and their absolute values were calculated. The signals were then segmented using a 100 ms window with a 50 ms overlap. Finally, the segmented EMG signals were concatenated across channels (Fig. \ref{fig:emg_processing}). These segmented EMG signals were acquired at 20 Hz intervals to match the sampling rate of the sensorless motion capture system and were fed continuously into the DNN model. This process ensured that the temporal length of the DNN input remained constant, while the number of channels varied according to the EMG length. We used the \texttt{butter} function and \texttt{sosfilt} function from \texttt{scipy.signal} in Python for implementation.

\subsection{Motion Labels Processing}
Motion labels (rest, flexion, and extension) were obtained from the elbow joint's angular velocity and angular acceleration. As an initial verification, we decided to predict a simple motion. Detailed information on this process is provided in the appendix.

\subsection{DNN Model}
\begin{figure*}[htbp]
\centering
\includegraphics[width=\textwidth]{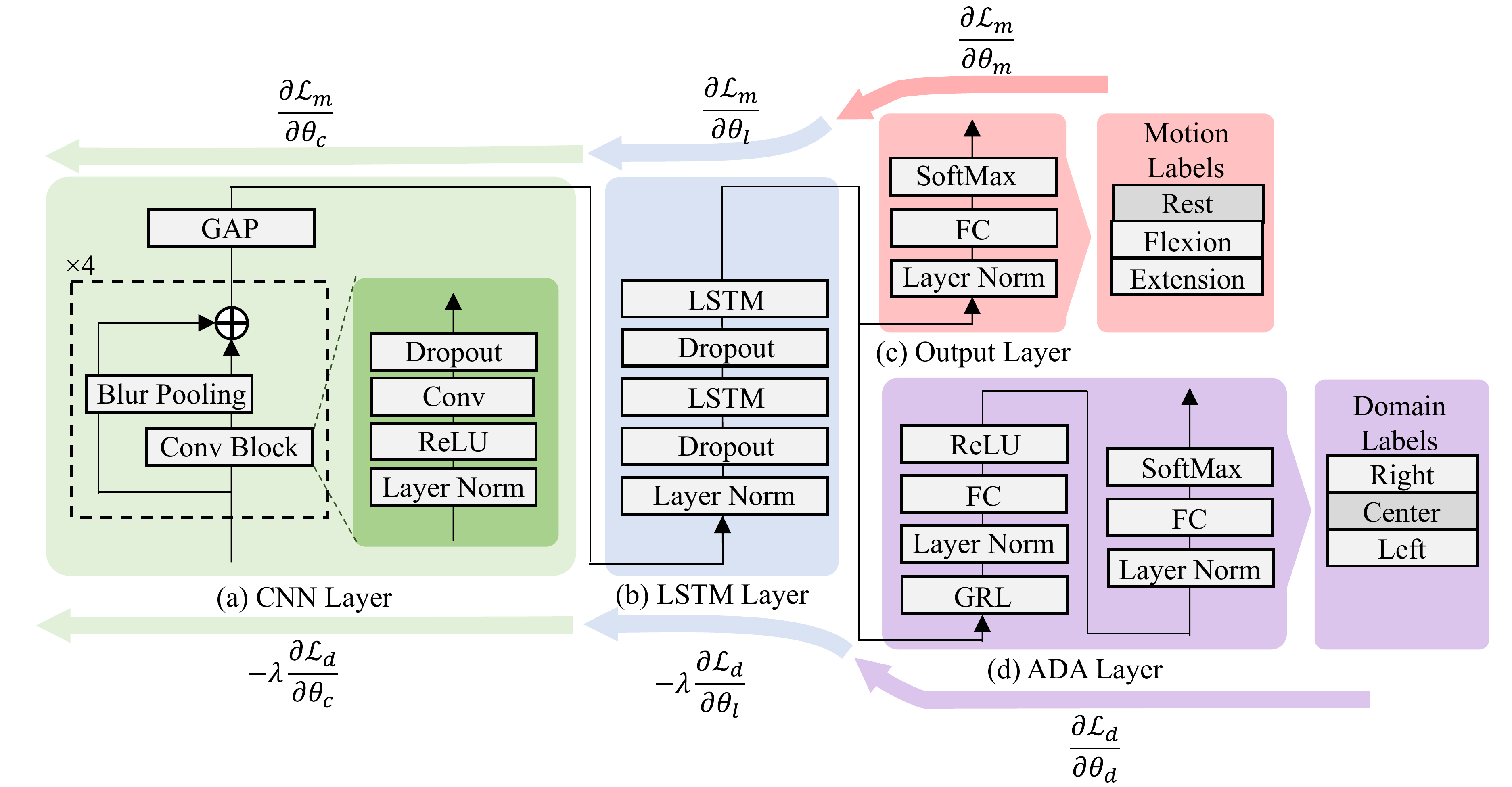}
\caption{The DNN model: (a) CNN Layer to extract temporal and channel-wise features, (b) LSTM Layer to extract temporal-related features, (c) Output Layer to predict motion labels, (d) ADA Layer to predict electrode positions: left, center, and right (domain labels) for adversarial domain adaptation. $\mathcal{L}_{m}$ is lost function for motion classifier, $\mathcal{L}_{d}$ is lost function for domain classifier, and $\theta_\text{c}$, $\theta_\text{l}$, $\theta_\text{m}$, $\theta_\text{d}$ are the weight parameters for CNN, LSTM, Output, and ADA Layers, and $\lambda$ is a parameter to control the gradient reversal strength.}
\label{fig:DNN_model}
\end{figure*}

We employed a CNN-LSTM model \cite{cnn-lstm1} as our DNN architecture (Fig. \ref{fig:DNN_model}), which is capable of extracting local spatiotemporal features via CNN layers and capturing temporal evolution via LSTM layers. This model predicts three classes: rest, flexion, and extension of the elbow. The model comprised four components: a CNN layer, an LSTM layer, an output layer, and an ADA layer. The CNN layer consists of four convolutional blocks, followed by Blur Pooling \cite{blur-pooling} and Global Average Pooling (GAP). Each convolutional block includes temporal layer normalization, a ReLU activation, a 1D convolution (with output channel size equal to the input channel size, kernel size: 3, stride: 1, and padding: 2), and dropout with rates of 0.1, 0.2, 0.3, and 0.4 for successive blocks. Blur Pooling (with a kernel size: 3, stride: 2, and padding: 1 applied only in the second block) is used to reduce aliasing effects during downsampling. The LSTM layer applies layer normalization, followed by two LSTM modules (with input and hidden sizes equal to the number of input channels) and dropout (rate: 0.1). The output layer includes layer normalization, a fully connected module (with output size equal to the number of input channels), and a SoftMax function. When adversarial domain adaptation is performed, ADA layer is activated. The ADA layer includes a Gradient Reversal Layer (GRL, with $\lambda=1.0$; a parameter to control the gradient reversal strength), two layer normalization layers, two fully connected modules (with output size equal to the number of input channels), a ReLU activation, and a SoftMax function. It is used solely to predict the electrode positions: left, center, and right. Focal Loss \cite{focal-loss} is used as the loss function in \eqref{eq:focalloss} since it facilitates training with unbalanced class labels.
\begin{equation}
\begin{gathered}
\mathcal{L}^\text{F}(\mathbf{y}, \hat{\mathbf{y}}) = \sum_{n=1}^{N} \alpha_l \, (1-p_n)^\gamma \, \mathcal{L}^\text{CE}(y_n, \hat{y}_n),\\[2mm]
\alpha_l = \frac{\operatorname{count}(\mathbf{y})_l}{\operatorname{count}(\mathbf{y})},\\[2mm]
p_n = \exp\Bigl(-\mathcal{L}^\text{CE}(y_n,\hat{y}_n)\Bigr).
\end{gathered}
\label{eq:focalloss}
\end{equation}
Here, $\alpha_l$ denotes the label rate for label $l$, $\gamma$ is a parameter that adjusts $(1-p_n)$, $\mathcal{L}^\text{CE}(y_n, \hat{y}_n)$ represents the cross-entropy loss between the $n^\text{th}$ true label $y_n$ and the $n^\text{th}$ predicted label $\hat{y}_n$, and $\operatorname{count}(\cdot)$ is a count function that returns the number of elements in its argument; specifically, $\operatorname{count}(\mathbf{y})_l$ denotes the number of true labels equal to $l$, and $\operatorname{count}(\mathbf{y})$ denotes the total number of labels. The DNN model was implemented using PyTorch 2.4.1 and PyTorch Lightning 2.4.0.

\subsection{Comparison Methods}
In this study, we first compared the classification performance achieved using SWN as a normalization method against other approaches, namely transfer learning (TL), adversarial domain adaptation (ADA), and a mixture of multiple electrode positions (MIX), as described in Section \ref{subsec:method_comparison}. Furthermore, since SWN is a normalization method that can be combined with other techniques, we also evaluated its integration with TL, ADA, and the MIX approach.

\subsubsection{Sliding-Window Normalization (SWN)}
Sliding-window normalization (SWN) is a real-time normalization technique that combines sliding-window processing with z-score normalization and is applied prior to feature extraction (Fig. \ref{fig:emg_processing}). It was originally proposed to reduce individual differences in EMG signals among subjects \cite{mypaper}. Previous study \cite{mypaper} has reported that SWN generally improves classification accuracy for both subject-specific and cross-subject models, although it does not necessarily reduce individual differences in EMG amplitude. We hypothesize that one reason for the improvement in classification performance is that SWN reduces differences among channels by aligning the EMG amplitude. For electrode shifts, we assumed that the EMG amplitude varies only slightly when shifted by a few centimeters, and that amplitude alignment improves classification performance. The SWN processing is defined in Eq. \eqref{eq:swn}:
\begin{equation}
y_{t,\, n-t+L^{\text{norm}}} = \frac{x_n - m_t}{s_t} \quad \bigl(t - L^{\text{norm}} < n \le t\bigr)
\label{eq:swn}
\end{equation}
Here, $t$ represents the current discrete time, $L^{\text{norm}}$ is the sliding window length, $n$ is the discrete-time index within the sliding window, $x_n$ denotes the $n^{\text{th}}$ processed EMG value, and $y_{t,\,n-t+L^{\text{norm}}}$ is the EMG signal after applying the SWN corresponding to the ${n-t+L^{\text{norm}}}^{\text{th}}$ element at time $t$. The parameters $m_t$ and $s_t$ represent the mean and standard deviation computed over the sliding window at time $t$, respectively. The \texttt{mean} and \texttt{std} functions in numpy were used in Python to compute these statistics.

\subsubsection{Vanilla}
The term “Vanilla” denotes to the plain model without any enhancements from TL, MIX, or ADA. In this case, the model comprises only the unmodified CNN-LSTM.

\subsubsection{Transfer Learning (TL)}
Transfer learning is a technique that leverages knowledge from a pre-trained model by initially training on a large source domain dataset, and then freezing the feature extraction layers while re-training the output layers with limited amount of target domain data. We selected transfer learning because it is one of the most effective methods for improving classification performance when adapting to shifted electrode data \cite{electrode-shift_tl1_mix_ref1}. In this study, we froze only the CNN layer and retrained the LSTM and output layers, as the time-related features captured by the LSTM may be affected by changes in electrode position.

\subsubsection{Adversarial Domain Adaptation (ADA)}
Adversarial domain adaptation (ADA) employs two output layers: one for class label classification and another for domain label classification. Typically, ADA is trained in an unsupervised manner using source domain data with class labels and target domain data without class labels, using a domain classifier that distinguishes between source and target domains \cite{recalibration_ada_electrode-shift_daily-effects}. However, in this study, we performed supervised adversarial multi-domain adaptation \cite{ada_ref}. In our approach, ADA was trained on mixed data including all three electrode positions (left, center, and right), and the domain classifier was designed to predict the specific electrode position rather than simply distinguishing between source and target domains. This strategy enables the ADA model to capture the diversity present in multi-condition data. We selected ADA for comparison with the proposed SWN because it, like SWN, does not require re-training when the electrodes are shifted.

\subsubsection{Mixture of Multiple Electrode Positions (MIX)}
The MIX approach is similar to ADA, but it does not employ a multi-domain classifier. In this method, models are trained using mixed data from all three electrode positions (left, center, and right) \cite{electrode-shift_tl1_mix_ref1}. We adopted MIX for comparison with ADA and SWN, as it does not require re-training when the electrodes are shifted.

\subsection{Training and Evaluation Criteria}

\subsubsection{Training, Tuning, and Testing Data}
The training, tuning, and testing datasets were defined according to the DNN strategy, as detailed in Table \ref{tab:evaluation_settings}. In the TL and Vanilla models, 70\% of the data from one electrode position were randomly selected for training. For TL, an additional 20\% of the data from the same electrode position were randomly used as tuning data so that both the tuning and testing data originated from the same electrode position. For the ADA and MIX models, the training data consisted of 20\% of the data from each of the three electrode positions, randomly selected. Meanwhile, 30\% of the data from one electrode position were randomly selected as the testing data. Notably, in the TL and Vanilla models,despite the difference between the electrode positions used for training and testing, the same testing data were consistently employed, ensuring a fair evaluation. The training epochs, batch size, time length for training data and optimizer settings were identical across TL, MIX, ADA, and Vanilla models: 40 training epochs, a batch size of 128, time length for training data of 20 s, and the Adam optimizer (learning rate: $1.0\times10^{-3}$, $\beta_1=0.9$, and $\beta_2=0.999$). In addition, for TL, we performed 10 re-training epochs for one of the different electrode positions.

\begin{table*}[htbp]
\setlength{\extrarowheight}{2mm}
\centering
\caption{The DNN strategy and various data settings. ADA and MIX do not conduct transfer learning.}
\begin{tabularx}{\textwidth}{|l|X|X|X|l|l|}
\hline
DNN strategy & Training data & Tuning data & Testing data \\[2mm]
\hline
TL, Vanilla  & 70\% of the data acquired from one electrode position             & 20\% of the data acquired from the electrode position employed for testing & 30\% of the data acquired from one of the different electrode positions from those used for training \\[2mm]
ADA, MIX     & 20\% of the data acquired from each of the three electrode positions          & -   & 30\% of the data acquired from one electrode position \\[2mm]
BASELINE & Same as TL                                                             & -   & 30\% of the data acquired from the electrode position employed for training    \\[2mm]
\hline
\end{tabularx}
\label{tab:evaluation_settings}
\end{table*}

\subsubsection{Testing Method}
In this section, we describe the selection process for comparison results. For the TL and Vanilla models, training and testing data were obtained from different electrode positions (resulting in 6 combinations). For the MIX and ADA models, training data were collected from all three electrode positions, while testing data were selected from one electrode position (yielding 3 combinations). The best classification performance was determined by computing the mean classification accuracy for each electrode position combination and subject, and by selecting the maximum mean performance over the range of window length parameters. For the no-normalization (None) case, the optimal classification performance was selected based on the feature extraction window lengths (ranging from 200 to 1000 ms in 200 ms intervals). For SWN, the optimal classification performance was chosen among various combinations of normalization window lengths (200 to 1000 ms in 200 ms intervals) and feature extraction window lengths (200 to 1000 ms in 200 ms intervals).

\subsubsection{Evaluation Index}
We used the differential classification accuracy as an evaluation metric to assess the reduction in the impact of electrode shift on classification performance. As defined in Eq. \eqref{eq:difference_accuracy}, this metric is calculated for each subject and each combination of training and testing electrode positions. The differential classification accuracy is computed by subtracting the accuracy of the BASELINE (obtained using the same electrode position for training and testing data, and the same normalization method) from the accuracy achieved by each DNN strategy.
\begin{equation}
\begin{gathered}
y_{n,i,j} = x_{n,i,j} - x^\text{BASELINE}_{n,j,j}\\[2mm]
x = \frac{\text{Success of Predictions}}{\text{Success of Predictions} + \text{Failure of Predictions}}
\end{gathered}
\label{eq:difference_accuracy}
\end{equation}
Here, $n$ denotes the normalization method, $i$ refers to the training electrode position data, and $j$ corresponds to the tuning and testing electrode position data. The variable $y_{n,i,j}$ represents the differential classification accuracy using the training data from electrode position $i$ and tuning and testing data from electrode position $j$ under normalization method $n$. The term $x_{n,i,j}$ denotes the classification accuracy obtained using training data from electrode position $i$ and tuning and testing data from electrode position $j$, while $x^\text{BASELINE}_{n,j,j}$ represents the best classification accuracy obtained using the same electrode position for both training and testing based on the optimal window lengths for normalization and feature extraction (BASELINE as described in Table \ref{tab:evaluation_settings}).

Statistical significance was evaluated using the Wilcoxon rank-sum test with a significance level of $p<0.05$, applying the Bonferroni correction for multiple comparisons. We calculated the mean differential classification accuracy across electrode position combinations for each subject, and used these averages to test for statistical significance. The \texttt{ranksums} function from \texttt{scipy.stats} and the \texttt{multipletests} function from \texttt{statsmodels.sandbox.stats.multicomp} in Python were used for this purpose. Additionally, a two-way Scheirer-Ray-Hare test was conducted for between-group comparisons between SWN and the no-normalization condition across TL, ADA, and MIX using the \texttt{scheirerRayHare} function from \texttt{rcompanion} and the \texttt{Formula} function from \texttt{robjects} in \texttt{rpy2} (R-4.4.3).

\section{Results}
We examined the impact of varying window lengths for normalization and feature extraction (Section \ref{subsec:Efects_of_Window_Lengths}), and then evaluated the performance of SWN and its integration with DNN strategies (Sections \ref{subsec:method_comparison} and \ref{subsec:DNN_comparison_with_swn}). \_SWN means with SWN, and \_None means no-normalization. All classification results were compared against a chance level of 33.3\% (three classes: rest, flexion, and extension).

\subsection{Effects of Window Lengths}
\label{subsec:Efects_of_Window_Lengths}
\begin{figure}[tbp]
\centering
\includegraphics[]{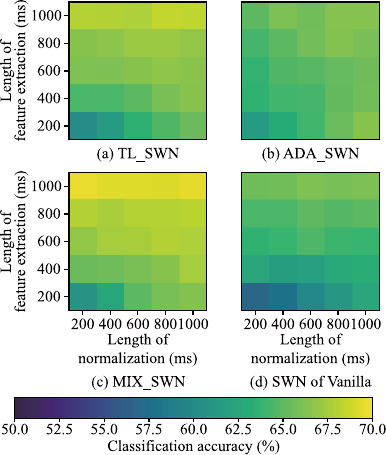}
\caption{The effects of window length on feature extraction with SWN. (a)-(d) indicate DNN strategies. The color indicates the classification accuracy}
\label{fig:acc_parameter_comparison_swn}
\end{figure}

\begin{figure}[tbp]
\centering
\includegraphics[]{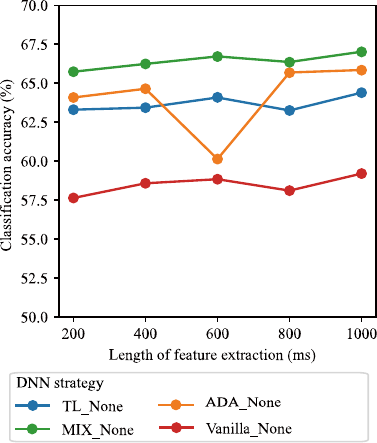}
\caption{The effects of window length on feature extraction without normalization.}
\label{fig:acc_parameter_comparison_none}
\end{figure}

We investigated how different window lengths for normalization and feature extraction affected classification accuracy across various DNN strategies and electrode positions. Fig. \ref{fig:acc_parameter_comparison_swn} and \ref{fig:acc_parameter_comparison_none} show the results when these window lengths are varied. For SWN (Fig. \ref{fig:acc_parameter_comparison_swn}), both the normalization and feature extraction window lengths were varied from 200 to 1000 ms in 200 ms increments. In the no-normalization (Fig. \ref{fig:acc_parameter_comparison_none}), only the feature extraction window lengths were varied over the same range.

In Fig. \ref{fig:acc_parameter_comparison_swn}(a)-(d), increasing the window lengths for both normalization and feature extraction consistently improved accuracy. In particular, Fig. \ref{fig:acc_parameter_comparison_swn}(a), (c) and (d) indicate that using a 1000 ms window for normalization enhances classification accuracy regardless of the feature extraction window length. The best performance in Fig. \ref{fig:acc_parameter_comparison_swn} was achieved by the MIX\_SWN, with a classification accuracy of 69.6\%. Based on these results, we recommend selecting a 1000 ms window for both normalization and feature extraction so that the window length maximizing accuracy can be chosen. The optimal window lengths were determined as follows: 600 ms for normalization and 1000 ms for feature extraction in SWN of Vanilla; 800 ms and 1000 ms in TL\_SWN; 1000 ms and 200 ms in ADA\_SWN; and 200 ms and 1000 ms in MIX\_SWN.

Similarly, in Fig. \ref{fig:acc_parameter_comparison_none}(a)-(d), increasing the feature extraction window length improved accuracy. The MIX\_None achieved the best result (67.0\% accuracy) with a 1000 ms window for feature extraction. This indicates that a 1000 ms window is optimal for feature extraction in the no-normalization condition. In Vanilla\_None, TL\_None, ADA\_None, MIX\_None, and BASELINE\_None, the optimal window length for feature extraction was 1000 ms.

\subsection{Comparison of Alternative Methods Against SWN}
\label{subsec:method_comparison}
\begin{figure}[tbp]
\centering
\includegraphics[]{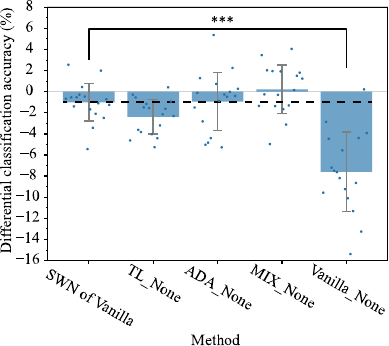}
\caption{Performance comparison between SWN and other compared methods. None indicates no-normalization. The blue dot markers in a blue bar indicate mean differential classification accuracy computed among electrode position combinations each subject, the error bar represents the standard deviation across subjects, the black solid line indicates the differential classification accuracy for SWN of Vanilla, and the gray line indicates 0\%. *** means $p<0.001$.}
\label{fig:diffacc_swn_vs_dnn_methods}
\end{figure}

To assess whether SWN improves classification accuracy under electrode shift conditions, we compared SWN of Vanilla with TL\_None, ADA\_None, MIX\_None, and Vanilla\_None (Fig. \ref{fig:diffacc_swn_vs_dnn_methods}). In SWN of Vanilla, the optimal BASELINE window lengths were 1000 ms for normalization and 600 ms for feature extraction. In contrast, for TL\_None, ADA\_None, MIX\_None, and Vanilla\_None, the optimal BASELINE window length for feature extraction was 1000 ms.

Fig. \ref{fig:diffacc_swn_vs_dnn_methods} shows that MIX\_None achieved the highest performance with a difference of 0.2\% in classification accuracy, which is 1.2\% higher than SWN of Vanilla (although not statistically significant, $p>0.05$). SWN of Vanilla tied with ADA\_None (both at -1.0\%) and outperformed TL\_None by 1.4\% ($p>0.05$) and Vanilla\_None by 6.6\% ($p<0.001$). These results indicate that SWN enhances classification accuracy across different electrode positions without requiring additional data, outperforming transfer learning (which depends on extra electrode position data) and achieving performance comparable to ADA and MIX.

\subsection{Comparison of DNN Methods with SWN Integration}
\label{subsec:DNN_comparison_with_swn}
\begin{figure}[tbp]
\centering
\includegraphics[]{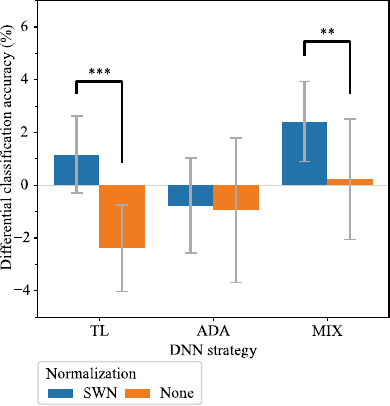}
\caption{Performance comparison between DNN strategies with SWN and without normalization. None indicates no-normalization. The error bar represents the standard deviation across subjects of the mean differential classification accuracy computed among electrode position combinations, the gray line indicates 0\%. ** means $p<0.01$, and *** means $p<0.001$.}
\label{fig:diffacc_swn_vs_none_of_dnn_methods}
\end{figure}

\begin{figure}[tbp]
\centering
\includegraphics[]{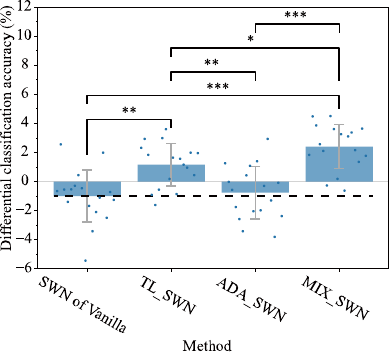}
\caption{Performance comparison among DNN strategies with SWN integration. The blue dot markers in a blue bar indicate mean differential classification accuracy computed among electrode position combinations each subject, the error bar represents the standard deviation across subjects, the black solid line indicates the result of SWN of Vanilla, and the gray line indicates 0\%. * means $p<0.05$, ** means $p<0.01$, and *** means $p<0.001$.}
\label{fig:diffacc_vs_dnn_methods_with_swn}
\end{figure}

In Section \ref{subsec:method_comparison}, we demonstrated that SWN improves classification accuracy when using the Vanilla model without additional electrode position data. Here, we investigated whether integrating SWN with various DNN strategies further enhances classification accuracy. Fig. \ref{fig:diffacc_swn_vs_none_of_dnn_methods} compares DNN strategies with SWN against their no-normalized counterparts. Furthermore, Fig. \ref{fig:diffacc_vs_dnn_methods_with_swn} compares SWN of Vanilla with other DNN strategies integrated with SWN. In SWN integrations, the optimal BASELINE window lengths were identical to those obtained for SWN of Vanilla in Section \ref{subsec:method_comparison}, whereas in the no-normalized methods, the optimal window length for feature extraction remained as described in Section \ref{subsec:method_comparison}.

Fig. \ref{fig:diffacc_swn_vs_none_of_dnn_methods} shows that applying SWN significantly improved the difference in classification accuracy for TL and MIX. A two-way Scheirer-Ray-Hare test revealed a highly significant difference between the SWN and no-normalization ($p=8.70\times10^{-10}$), indicating that the effect of SWN is consistent across methods. Specifically, TL\_SWN achieved a 1.2\% difference, which is 3.6\% higher than TL\_None (-2.4\%, $p<0.001$), and MIX\_SWN achieved a 2.4\% difference, which is 2.2\% higher than MIX\_None (0.2\%, $p<0.01$). In contrast, ADA\_SWN (-0.8\%) was not significantly different from ADA\_None (-0.9\%, $p>0.05$). These results indicate that integrating SWN with DNN strategies further enhances classification accuracy under electrode shift conditions.

Fig. \ref{fig:diffacc_vs_dnn_methods_with_swn} clearly shows that SWN integrations  significantly improve classification accuracy compared to SWN of Vanilla. Among the SWN-integrated methods, MIX\_SWN achieved the highest improvement at 2.4\%, followed by TL\_SWN with a 1.2\% increase. Moreover, MIX\_SWN outperformed TL\_SWN by 1.3\% ($p<0.05$), and compared to SWN of Vanilla, TL\_SWN and MIX\_SWN showed improvements of 2.2\% and 3.4\%, respectively. Both TL\_SWN and MIX\_SWN also surpassed the BASELINE (0\%) by 1.3\% and 2.4\%, respectively.

\section{Discussion}
In this study, we applied sliding-window normalization (SWN) to mitigate the reduction in classification accuracy caused by electrode shift. We succeeded in improving the differential classification accuracy from -7.6\% to -1.0\%—an improvement of 6.6\%—by applying SWN (Fig. \ref{fig:diffacc_swn_vs_dnn_methods}). These results demonstrate the effectiveness of SWN.

In this section, we discuss (1) the performance of SWN of Vanilla compared with alternative DNN strategies without normalization (Section \ref{subsec:peformance_of_proposed_swn}), (2) the performance of DNN strategies integrated with SWN (Section \ref{subsec:dnn_methods_with_proposed_swn_integration}), and (3) the selection of parameters for SWN (Section \ref{subsec:parameters_selection_for_proposed_swn}).

\subsection{Performance of SWN}
\label{subsec:peformance_of_proposed_swn}
SWN mitigated the reduction in classification accuracy due to electrode shift; in particular, the SWN of Vanilla (-1.0\%) achieved a 6.6\% improvement in differential classification accuracy compared to Vanilla\_None (-7.6\%) (Fig. \ref{fig:diffacc_swn_vs_dnn_methods} in Section \ref{subsec:method_comparison}). Moreover, its performance is nearly equivalent to ADA\_None (-0.9\%) and 1.4\% higher than TL\_None (-2.4\%). However, the SWN of Vanilla performed 1.2\% lower than MIX\_None (0.2\%). Although the performance of the SWN of Vanilla is not the highest, it is noteworthy that SWN enhances results using data from only a single electrode position, whereas other DNN strategies require data from multiple electrode positions. This incremental improvement is attributable to the fact that our method does not require any additional data, thereby significantly reducing the calibration and retraining efforts and costs in practical applications, which is extremely advantageous for real-world use. Furthermore, as our method can be readily integrated with the latest DNN architectures and applied to more complex multi-class motion prediction problems, further performance enhancements can be anticipated, suggesting a promising avenue for future research and practical deployment.

Surprisingly, the best performance was achieved with MIX\_None (Fig. \ref{fig:diffacc_swn_vs_dnn_methods}), which simply combines data from multiple electrode positions without employing adversarial domain adaptation or transfer learning. This higher performance compared to BASELINE\_None (Table \ref{tab:evaluation_settings}) is likely due to the model being trained on a mixture of data from multiple electrode positions, enabling it to adapt to a variety of environments. This phenomenon was also observed in a previous study on cross-subject generalization (i.e., a model trained on data from other subjects) \cite{mypaper}, where increasing the number of subjects improved classification accuracy. Although we initially expected ADA\_None to outperform MIX\_None, MIX\_None achieved the best performance. These findings suggest that training with a mixture of multiple conditions is superior to specialized DNN strategies—especially when electrode positions are shifted by approximately 2 cm and multiple datasets with motion labels are available. A shift of approximately 2 cm results in only minor changes in the amplitude and local features of the EMG signals, while the overall pattern remains largely unchanged, making it an ideal condition for capturing and learning subtle variations. In contrast, a previous study \cite{electrode-shift_tl1_mix_ref1} (ten-class scenario) reported that transfer learning (TL) achieved a 6\% classification error—1\% lower than the 7\% error obtained with the MIX approach—which is the opposite of our findings. We attribute this discrepancy to a data imbalance, as 75\% of the training data came from no-shifted electrodes while only 25\% came from shifted electrodes. In that study, the data from the no-shifted condition were approximately three times more abundant than those from the shifted condition, likely causing the model to become overly adapted to the no-shifted state and resulting in higher accuracy for TL than MIX. In contrast, our study—where the data were more evenly balanced—demonstrated that the MIX approach outperformed TL. Thus, the composition of the training data is a crucial factor in selecting a strategy to mitigate performance degradation. Moreover, the MIX approach is the simplest, as it merely involves combining data from multiple electrode positions.

\subsection{DNN strategies with SWN Integration}
\label{subsec:dnn_methods_with_proposed_swn_integration}
The TL and MIX models integrated with SWN show improved performance compared with their no-normalized counterparts. Specifically, TL\_SWN exhibits a 3.6\% higher differential classification accuracy than TL\_None, and MIX\_SWN achieves the best performance with a 2.4\% improvement—2.2\% higher than MIX\_None—as shown in Section \ref{subsec:peformance_of_proposed_swn}. In contrast, ADA\_SWN improved by only 0.2\%, showing little difference compared with ADA\_None. These results indicate that TL and MIX exhibit a synergistic effect when integrated with SWN. Furthermore, as noted earlier, training with a mixture of multiple conditions is superior to specialized DNN strategy, particularly when electrode positions are shifted by approximately 2 cm and multiple datasets with motion labels are available. Moreover, we believe that the electrode shift issue is effectively addressed, as both TL\_SWN and MIX\_SWN outperformed models trained and tested on data from the same electrode position. To our knowledge, no previous study has demonstrated that a normalization approach can mitigate the electrode shift issue to the extent observed in our work. In addition, no prior study has combined multiple methods to address electrode shift, nor has any prior study achieved performance that exceeds the baseline. Our results indicate that integrating SWN with DNN strategies (TL or MIX)—effectively mitigates the electrode shift issue, outperforming even models trained and tested on data from a single electrode position.

\subsection{Parameters Selection for Proposed SWN}
\label{subsec:parameters_selection_for_proposed_swn}
Fig. \ref{fig:acc_parameter_comparison_swn} and \ref{fig:acc_parameter_comparison_none} indicate that longer window lengths for both normalization and feature extraction (between 200 and 1000 ms) yield higher classification accuracy. However, extending the window length for normalization appears to be more effective than extending that for feature extraction. In particular, as shown in Fig. \ref{fig:acc_parameter_comparison_swn}(c), using a 1000 ms window for normalization results in nearly identical classification accuracy regardless of the feature extraction window length for the MIX method. Therefore, to reduce computation time, we recommend using a 1000 ms window for normalization and a 200 ms window for feature extraction.

\section{Conclusions}
In this paper, we addressed the reduction in classification accuracy due to electrode shift by applying sliding-window normalization (SWN). We evaluated performance using differential classification accuracy between each DNN strategy (trained on one electrode position and tested on another) and the BASELINE (trained and tested on the same electrode position). SWN (Vanilla CNN-LSTM with SWN) improved differential accuracy by 6.6\% compared to no SWN, outperforming TL\_None and ADA\_None. Although MIX\_None achieved a 1.2\% higher performance than SWN of Vanilla, SWN's advantage is that it requires no additional electrode data, significantly reducing calibration and retraining efforts. Moreover, combining SWN with recent strategies yielded the best result, with MIX\_SWN achieving a 2.4\% improvement over BASELINE\_SWN, clearly demonstrating that SWN effectively mitigates electrode shift and enhances classification accuracy.

Future work will pursue several avenues to enhance our approach. First, we will integrate data from multiple subjects \cite{mypaper} and examine the impact of daily variations \cite{recalibration_ada_electrode-shift_daily-effects, recalibration_daily-effects2, recalibration_daily-effects3} on EMG amplitude and model performance, aiming to enhance generalizability not only to electrode position variations but also to inter-subject differences and daily fluctuations, thereby ensuring long-term stability in real-world conditions. Second, we plan to expand classification outputs by incorporating additional movement types—forearm pronation/supination and shoulder joint flexion/extension—to enable comprehensive 3-dimensional motion prediction \cite{multi-est1, multi-est2}. Third, previous studies \cite{amputee1, amputee2} have shown that transfer learning improves classification accuracy, suggesting it is expected that our approach may also achieve improved predictive performance when applied to amputee data. Finally, we will explore applying SWN to regression tasks for continuous output prediction.

\appendix
\section*{Appendix}
\section{Comparison of Alternative Methods Against the SWN}
\begin{figure*}[htbp]
\centering
\includegraphics[width=\textwidth]{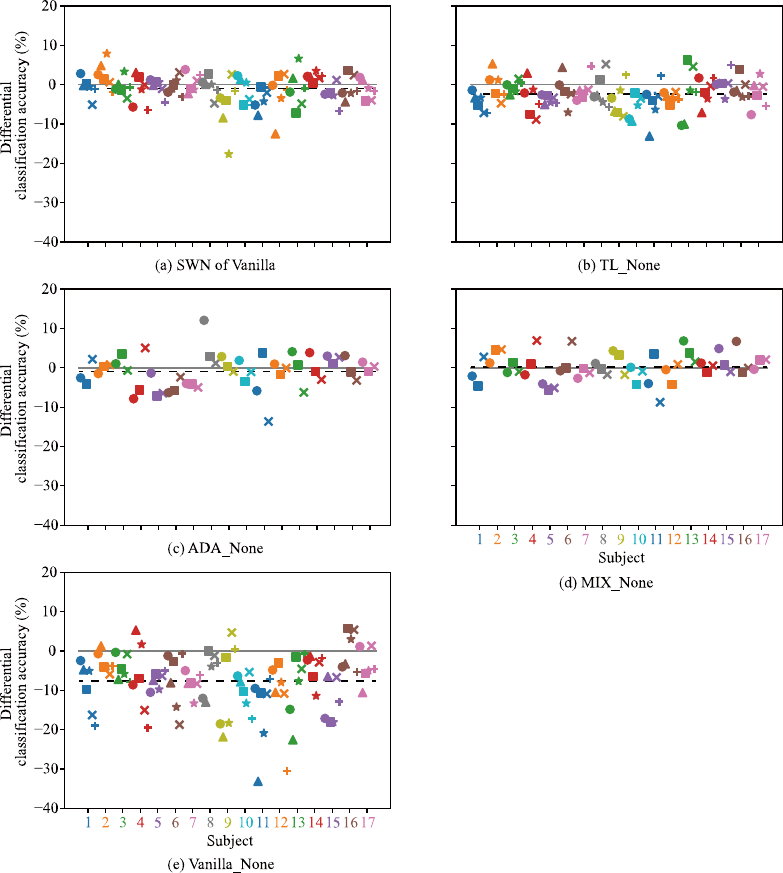}
\caption{Performance comparison each subject between SWN and alternative methods. (a)-(e) indicate method name. The black solid lines indicate mean differential classification accuracy among electrode position combinations and subjects each method, and the gray line indicates 0\%.}
\label{fig:diffacc_swn_vs_dnn_methods_scatter}
\end{figure*}

In the Section \ref{subsec:method_comparison}, we compared the performance between SWN of Vanilla and alternative methods without normalization. In this section, we investigate the difference classification accuracy each subject between SWN of Vanilla and TL\_None, ADA\_None, MIX\_None, and Vanilla\_None (Fig. \ref{fig:diffacc_swn_vs_dnn_methods_scatter}). We compared the best performance results each method in the window lengths for normalization and feature extraction, like Section \ref{subsec:method_comparison}. The window lengths for normalization and feature extraction are changed in the range of 200-1000 ms in 200 ms increments. In the Fig. \ref{fig:diffacc_swn_vs_dnn_methods_scatter}, the black solid line shows mean and the gray line shows 0\%.

From Fig. \ref{fig:diffacc_swn_vs_dnn_methods_scatter}, there are same performance behavior  among subjects in the A-E. The difference class accuracy was best in the order of MIX\_None, SWN of Vanilla, like in Section \ref{subsec:method_comparison}. 

\section{Comparison of DNN methods with the SWN Integration}
\begin{figure*}[htbp]
\centering
\includegraphics[width=\textwidth]{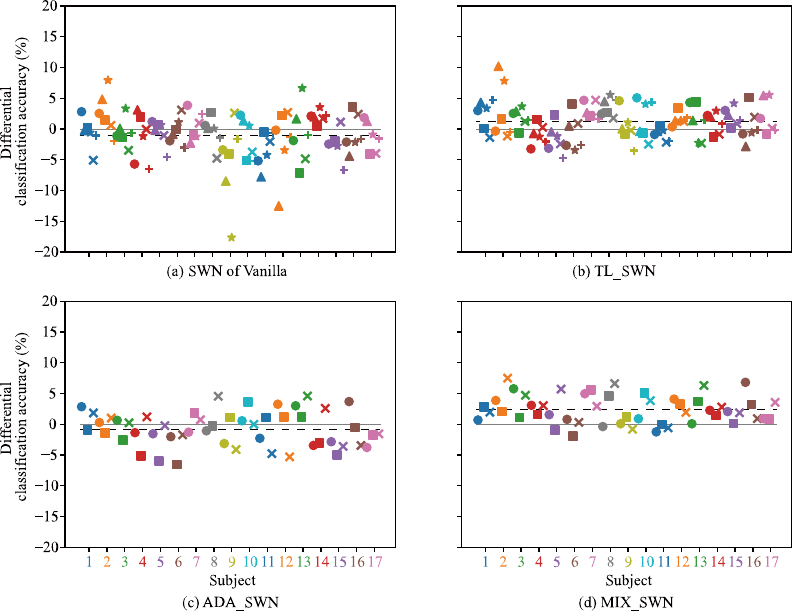}
\caption{Performance comparison each subject between DNN methods with SWN and without normalization. (a)-(d) indicate DNN method name. The black solid lines indicate mean differential classification accuracy among electrode position combinations and subjects each method, and the gray line indicates 0\%.}
\label{fig:diffacc_vs_dnn_methods_with_swn_scattter}
\end{figure*}

In the Section \ref{subsec:DNN_comparison_with_swn}, we compared the performance among DNN methods with SWN. In this section, we investigate the difference classification accuracy each subject among DNN methods with SWN (Fig. \ref{fig:diffacc_vs_dnn_methods_with_swn_scattter}). In Fig. \ref{fig:diffacc_vs_dnn_methods_with_swn_scattter}, the black solid line shows mean and the gray line shows 0\%.

From Fig. \ref{fig:diffacc_vs_dnn_methods_with_swn_scattter}, there are same performance behavior  among subjects in the A-D. The difference class accuracy was best in the order of MIX\_SWN, TL\_SWN, like in Section \ref{subsec:DNN_comparison_with_swn}. 

\section{Sensorless Motion Capture System}
\label{sec: sensorless_motion_capture_system}
We use DeepLabCut-Live \cite{dlc-live} and anipose \cite{anipose} to capture human body part positions. We describe the how to get the elbow and shoulder joint angles from images, initializing multi-view triangulation, and initializing the rotating matrix and shifting vector.

\subsection{Discription of Sensorless Motion Capture System}
We show the progress in getting the elbow and shoulder joint angles from images.
\begin{enumerate}
\item Calibrating for 6$\text{)}$ Multi-view triangulation and 7$\text{)}$ Coordinate transformation with a $5\times5$ checkerboard.
\item Capturing three images (20 Hz) by OpenCV (C++) with a video capture board.
\item Resizing images $1920\times1080$ to $640\times360$.
\item Sending resized images from C++ to Python.
\item Estimating body part positions in each image by DeepLabCut-Live.
\item Multi-view triangulation by anipose.
\item Coordinate transformation by \eqref{eq:coordinate_transformation} from the 3D triangulation space to the 3D human space. In the 3D triangulation space:$\text{D}^\text{T}$, the three-dimensional coordinate system is formed based on the camera's orientation and is therefore not necessarily horizontal with respect to the ground. To address this, we rotate the coordinate system and transform it into a three-dimensional coordinate system that is aligned with the ground: $\text{D}^\text{H}$, making it more convenient for human interpretation.
\begin{equation}
\bm{p}^\text{H} \;=\; \text{R}^\text{xyz}\,\bm{p}^\text{T} \;+\; \bm{t}
\label{eq:coordinate_transformation}
\end{equation}
\item Sending body part positions from Python to C++.
\item Transforming wrist, elbow, and shoulder positions to elbow and shoulder joint angles by \eqref{eq:task2joint}.
\end{enumerate}

\subsection{Initializing Multi-View Triangulation}
In order to perform triangulation, the transformation parameters must be initialized. We show procedure that initializes multi-view triangulation (anipose).
\begin{enumerate}
\item Saving three synchronized videos showing a checkerboard. The checkerboard must be photographed at several angles and positions. This checkerboard was made by \texttt{cv2.aruco.CharucoBoard} class in OpenCV (Python), $5\times5$ grids, 0.17 m square length, 0.15 m marker length, 4 marker bits, and 50 marker kinds.
\item Initializing by anipose using saved videos.
\end{enumerate}

\subsection{Initializing the Rotating Matrix and Shifting Vector}
\begin{figure*}[htbp]
\centering
\includegraphics[width=\textwidth]{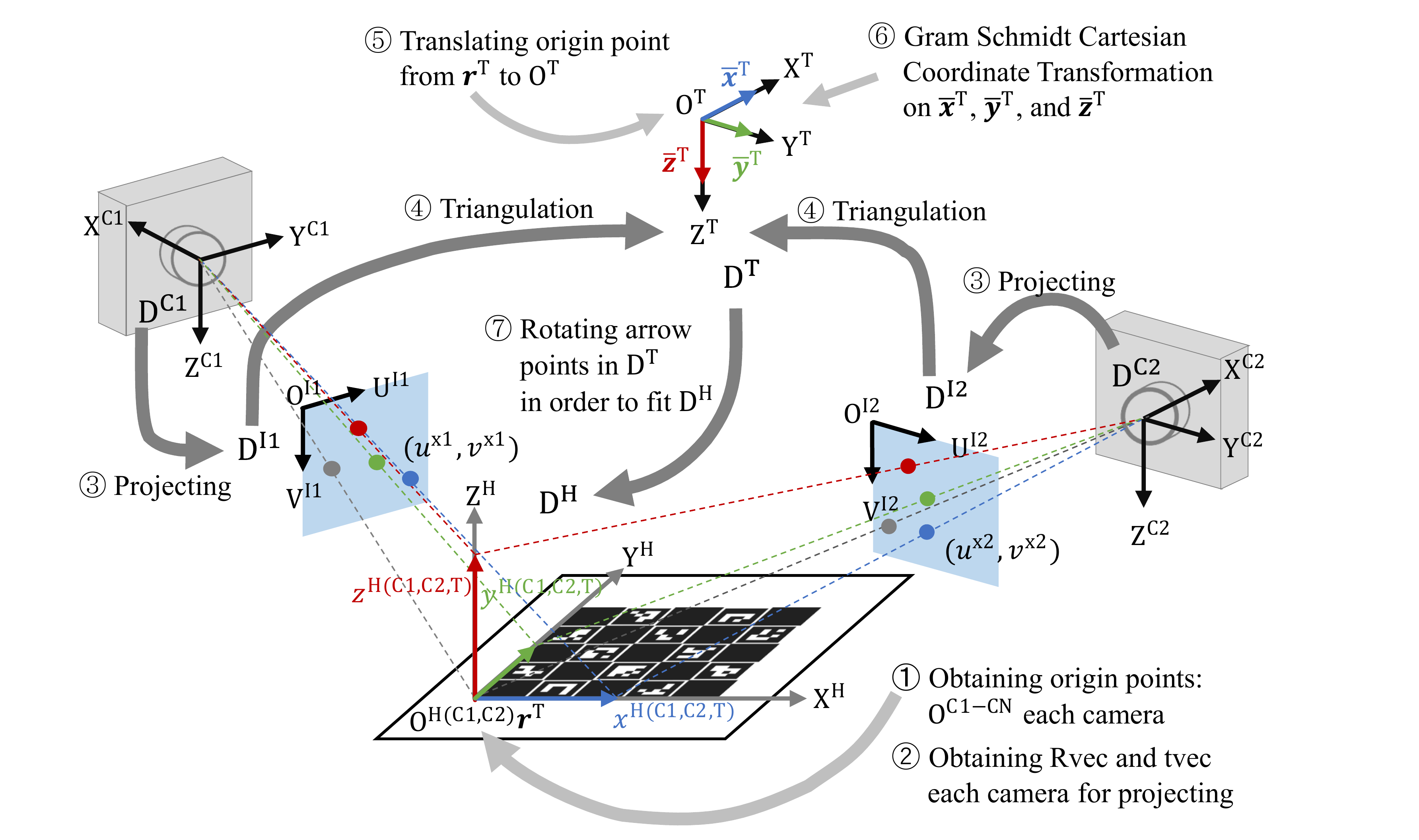}
\caption{An explanation of the method for transforming arrow points from the 3D camera space into the 3D human space. $\text{D}^\text{H}$ means a 3D space of the human, $\text{D}^\text{C}$ means a 3D space of the camera, $\text{D}^\text{I}$ means a 2D space of the image, and $\text{D}^\text{T}$ means a 3D space after triangulation. the circled numbers regard to numbers in \textbf{Initializing the rotating matrix and shifting vector}. To obtain rotation matrix and translation vector to transform any points in $\text{D}^\text{T}$ ioto $\text{D}^\text{H}$, the blue, green, and red arrow points and a origin point in $\text{D}^\text{C}$ are obtained, and transformed into in order $\text{D}^\text{I}$, $\text{D}^\text{T}$. The translation vector is obtained by translating a origin point from $\text{O}^\text{T}$ to $\text{O}^\text{H}$. Further, the rotation matrix is obtained by rotating three arrow points individually.}
\label{fig:coordinates_transformation}
\end{figure*}

We describe the method for obtaining the rotation matrix: $\text{R}^\text{xyz}$ and translation vector: $\bm{t}$ used to transform the 3D triangulation space:$\text{D}^\text{T}$ into a coordinate system that is horizontally aligned with the ground: $\text{D}^\text{H}$. Fig. \ref{fig:coordinates_transformation} indicates how to get the $\text{R}^\text{xyz}$ and $\bm{t}$. They are got by allow points ($\bm{x}^{\text{C1\textasciitilde CN}},\ \bm{y}^{\text{C1\textasciitilde CN}},\ \bm{z}^{\text{C1\textasciitilde CN}}$) and origins ($\mathbf{O}^{\text{C1\textasciitilde CN}}$) shown in the images. Here, $\text{N}$ is the number of the cameras.
\begin{enumerate}
\item Getting the origin position of the checkerboard in the image for each camera.  
We used \texttt{detectBoard} function in the \texttt{cv2.aruco.CharucoDetector} class in the OpenCV in Python.
\item Getting $\bm{Rvec}$ and $\bm{tvec}$ to transform between 
$\text{D}^{\text{I}}$ (2D image space) and $\text{D}^{\text{C}}$ (3D camera space) for the origin of the checkerboard for each camera.  
We used \texttt{cv2.aruco.estimatePoseCharucoBoard} function in the OpenCV in Python.  
The $\bm{Rvec}$ is a vector to rotate the object coordinate to the camera coordinate,  
and the $\bm{tvec}$ is a vector to translate the origin of the object coordinate to the camera coordinate.
\item Projecting the arrow points: $\bm{x}^{\text{C1\textasciitilde CN}}$, $\bm{y}^{\text{C1\textasciitilde CN}}$, and $\bm{z}^{\text{C1\textasciitilde CN}}$, and the origin of the checkerboard: $\mathbf{O}^{\text{C1\textasciitilde CN}}$, form $\text{D}^{\text{C1\textasciitilde CN}}$ to $\text{D}^{\text{I1\textasciitilde IN}}$ by \texttt{cv2.projectPoints} function in OpenCV in Python.
We set $\bm{x}^{\text{C1\textasciitilde CN}}$, $\bm{y}^{\text{C1\textasciitilde CN}}$, and $\bm{z}^{\text{C1\textasciitilde CN}}$ as \dots
\begin{equation}
\begin{gathered}
\left[\bm{x}^{\text{C1\textasciitilde CN}},\ \bm{y}^{\text{C1\textasciitilde CN}},\ \bm{z}^{\text{C1\textasciitilde CN}},\ \mathbf{O}^{\text{C1\textasciitilde CN}}\right] = \\[2mm]
L^\text{a}
\begin{pmatrix}
0 & 1 & 0 & 0 \\
1 & 0 & 0 & 0 \\
0 & 0 & -1 & 0 
\end{pmatrix}
\end{gathered}
\label{eq:arrow_points_setting_in_DC}
\end{equation}
where, $L^\text{a}$ is the arrow length of the checkerboard that was set to 0.2 [m]. We use the cmtx and dist that were acquired by anipose to conduct \texttt{cv2.projectPoints} function. Here, the cmtx means camera matrix which is a $3\times3$ matrix representing the intrinsic camera parameters and the dist means the parameters for correcting image distortion caused by camera lenses.
\item Multi-view triangulation by anipose to transform allow points: $\bm{x}^\text{T}$, $\bm{y}^\text{T}$, and $\bm{z}^\text{T}$ and origin point of the checkerboard: $\bm{r}^\text{T}$.
\item Getting the arrow vectors and a shifting vector.
\begin{equation}
\left[\bm{\bar{x}}^\text{T},\ \bm{\bar{y}}^\text{T},\ \bm{\bar{z}}^\text{T}\right] = \left[\bm{x}^\text{T},\ \bm{y}^\text{T},\ \bm{z}^\text{T}\right] - \bm{r}^\text{T}
\label{eq:normalizing_arrow_position_in_DT}
\end{equation}
Where, $\bm{x}^\text{T}$, $\bm{y}^\text{T}$, and $\bm{z}^\text{T}$ indicates the arrow vectors. Further, the shifting vector is shown \eqref{eq:shifting_vector}.
\begin{equation}
\bm{t}=-\bm{r}^\text{T}
\label{eq:shifting_vector}
\end{equation}
\item Gram-Schmidt Cartesian Coordinate Transformation for arrow vectors. Transforming in the order $\bm{\bar{x}}^\text{T} - \bm{\bar{y}}^\text{T}$, $\bm{\bar{x}}^\text{T} - \bm{\bar{z}}^\text{T}$, $\bm{\bar{y}}^\text{T} - \bm{\bar{z}}^\text{T}$.
\item Getting coordinate rotation matrix. We find a coordinate rotation matrix, $\text{R}^\text{xyz}$, that satisfies \eqref{eq:transforming_DT_to_DH}.
\begin{equation}
\begin{gathered}
\begin{bmatrix}
\bm{\bar{x}}^\text{H} & \bm{\bar{y}}^\text{H} & \bm{\bar{z}}^\text{H}
\end{bmatrix}
=
\text{R}^\text{xyz}\,
\begin{bmatrix}
\bm{\bar{x}}^\text{T} & \bm{\bar{y}}^\text{T} & \bm{\bar{z}}^\text{T}
\end{bmatrix} \\[2mm]
\bm{\bar{x}}^\text{H} = {\begin{bmatrix} 1 & 0 & 0 \end{bmatrix}}^\text{T}\\[2mm]
\bm{\bar{y}}^\text{H} = {\begin{bmatrix} 0 & 1 & 0 \end{bmatrix}}^\text{T}, 
\quad \bm{\bar{z}}^\text{H} = {\begin{bmatrix} 0 & 0 & 1 \end{bmatrix}}^\text{T}
\end{gathered}
\label{eq:transforming_DT_to_DH}
\end{equation}
where, $\bm{\bar{x}}^\text{H}$, $\bm{\bar{y}}^\text{H}$, $\bm{\bar{z}}^\text{H}$ are normalized $\bm{x}^\text{H}$, $\bm{y}^\text{H}$, $\bm{z}^\text{H}$.

We rotate the vectors by the roll-pitch-yaw method as \eqref{eq:rotation_matrices}.
\begin{equation}
\begin{gathered}
\operatorname{Rot}(\theta_{\mathrm{x}}, \theta_{\mathrm{y}}, \theta_{\mathrm{z}})
= \operatorname{Yaw}(\theta_{\mathrm{z}})\,\operatorname{Pitch}(\theta_{\mathrm{y}})\,\operatorname{Roll}(\theta_{\mathrm{x}}) \\[2mm]
\operatorname{Roll}(\theta_{\mathrm{x}}) =
\begin{pmatrix}
1 & 0 & 0 \\
0 & \cos\theta_{\mathrm{x}} & -\sin\theta_{\mathrm{x}} \\
0 & \sin\theta_{\mathrm{x}} & \cos\theta_{\mathrm{x}}
\end{pmatrix} \\[2mm]
\operatorname{Pitch}(\theta_{\mathrm{y}}) =
\begin{pmatrix}
\cos\theta_{\mathrm{y}} & 0 & \sin\theta_{\mathrm{y}} \\
0 & 1 & 0 \\
-\sin\theta_{\mathrm{y}} & 0 & \cos\theta_{\mathrm{y}}
\end{pmatrix} \\[2mm]
\operatorname{Yaw}(\theta_{\mathrm{z}}) =
\begin{pmatrix}
\cos\theta_{\mathrm{z}} & -\sin\theta_{\mathrm{z}} & 0 \\
\sin\theta_{\mathrm{z}} & \cos\theta_{\mathrm{z}} & 0 \\
0 & 0 & 1
\end{pmatrix}
\end{gathered}
\label{eq:rotation_matrices}
\end{equation}
The coordinate transforming is conducted by \eqref{eq:Rxyz}.
\begin{equation}
\text{R}^{\text{xyz}} = {\text{R}^{\text{z}}}' \, \text{R}^{\text{y}} \, \text{R}^{\text{x}}
\label{eq:Rxyz}
\end{equation}
$\text{R}^\text{x}$ is gotten from \eqref{eq:rotation_x}, $\text{R}^\text{y}$ is gotten from \eqref{eq:rotation_y}, and ${\text{R}^\text{z}}'$ is gotten from \eqref{eq:rotation_z}.\\
\textbf{Rotating x axis}.
\begin{equation}
\begin{gathered}
\text{R}^{\text{x}} = \operatorname{Rot}(0, \theta_{\mathrm{y}}^1, \theta_{\mathrm{z}}^1) \\[2mm]
\theta_{\mathrm{y}}^1 = \operatorname{atan2d}\Bigl(\text{P}_{\mathrm{xz}}^1,\, \text{P}_{\mathrm{xx}}^1\Bigr) \\[2mm]
\theta_{\mathrm{z}}^1 =
\begin{cases}
\operatorname{atan2d}\!\Bigl(\text{P}_{\mathrm{xy}}^1,\, \dfrac{\text{P}_{\mathrm{xx}}^1}{\cos\theta_{\mathrm{y}}^1}\Bigr)\\[2mm] \hspace{4em} (-45^\circ < \theta_{\mathrm{y}}^1 < 45^\circ \text{ or } \\[2mm] \hspace{4em} 135^\circ < \theta_{\mathrm{y}}^1 < 225^\circ) \\[1ex]
\operatorname{atan2d}\!\Bigl(\text{P}_{\mathrm{xy}}^1,\, -\dfrac{\text{P}_{\mathrm{xz}}^1}{\sin\theta_{\mathrm{y}}^1}\Bigr) \hspace{1em} (\text{otherwise})
\end{cases} \\[2mm]
\text{P}^1 = \begin{bmatrix} \bar{x}^\text{T},\, \bar{y}^\text{T},\, \bar{z}^\text{T} \end{bmatrix}
\end{gathered}
\label{eq:rotation_x}
\end{equation}    

\vspace{\baselineskip}
\textbf{Rotating y axis}.
\begin{equation}
\begin{gathered}
\text{R}^{y} = \operatorname{Rot}(\theta_{\mathrm{x}}^2,\, 0,\, \theta_{\mathrm{z}}^2) \\[2mm]
\theta_{\mathrm{x}}^2 = \operatorname{atan2d}\Bigl(\text{P}_{\mathrm{yz}}^2,\, \text{P}_{\mathrm{yy}}^2\Bigr) \\[2mm]
\theta_{\mathrm{z}}^2 =
\begin{cases}
\operatorname{atan2d}\!\Bigl(\text{P}_{\mathrm{yx}}^2,\, \dfrac{\text{P}_{\mathrm{yy}}^2}{\cos\theta_{\mathrm{x}}^2}\Bigr) \\[2mm] \hspace{4em} (-45^\circ < \theta_{\mathrm{x}}^2 < 45^\circ \text{ or } \\[2mm] \hspace{4em} 135^\circ < \theta_{\mathrm{x}}^2 < 225^\circ) \\[1ex]
\operatorname{atan2d}\!\Bigl(\text{P}_{\mathrm{zz}}^2,\, -\dfrac{\text{P}_{\mathrm{yz}}^2}{\sin\theta_{\mathrm{x}}^2}\Bigr) \hspace{1em} (\text{otherwise})
\end{cases} \\[2mm]
\text{P}^2 = \text{R}^{\text{x}}\,\text{P}^1
\end{gathered}
\label{eq:rotation_y}
\end{equation}

\vspace{\baselineskip}
\textbf{Rotating z axis}.
\begin{equation}
\begin{gathered}
{\text{R}^\text{z}}' =
\begin{pmatrix}
1 & 0 & 0 \\
0 & 1 & 0 \\
0 & 0 & \mathrm{R}_{\mathrm{zz}}^{\mathrm{z}}
\end{pmatrix} \\[2mm]
\mathrm{R}^{\mathrm{z}} = \operatorname{Rot}(\theta_{\mathrm{x}}^3,\, \theta_{\mathrm{y}}^3,\, 0) \\[2mm]
\theta_{\mathrm{x}}^3 = \operatorname{atan2d}\Bigl(\mathrm{P}_{\mathrm{zx}}^3,\, \mathrm{P}_{\mathrm{zz}}^3\Bigr) \\[2mm]
\theta_{\mathrm{y}}^3 =
\begin{cases}
\operatorname{atan2d}\!\Bigl(\mathrm{P}_{\mathrm{zx}}^3,\, \dfrac{\mathrm{P}_{\mathrm{zz}}^3}{\cos\theta_{\mathrm{x}}^2}\Bigr) \\[2mm] \hspace{4em} (-45^\circ < \theta_{\mathrm{x}}^3 < 45^\circ \text{ or } \\[2mm] \hspace{4em} 135^\circ < \theta_{\mathrm{x}}^3 < 225^\circ) \\[1ex]
\operatorname{atan2d}\!\Bigl(\mathrm{P}_{\mathrm{zz}}^3,\, -\dfrac{\mathrm{P}_{\mathrm{zy}}^3}{\sin\theta_{\mathrm{x}}^3}\Bigr) \hspace{1em} (\text{otherwise})
\end{cases} \\[2mm]
\mathrm{P}^3 = \mathrm{R}^{\mathrm{y}}\,\mathrm{P}^2
\end{gathered}
\label{eq:rotation_z}
\end{equation}
\end{enumerate}

\section{Generated Tasks}
\begin{figure*}[htbp]
\centering
\includegraphics[width=\textwidth]{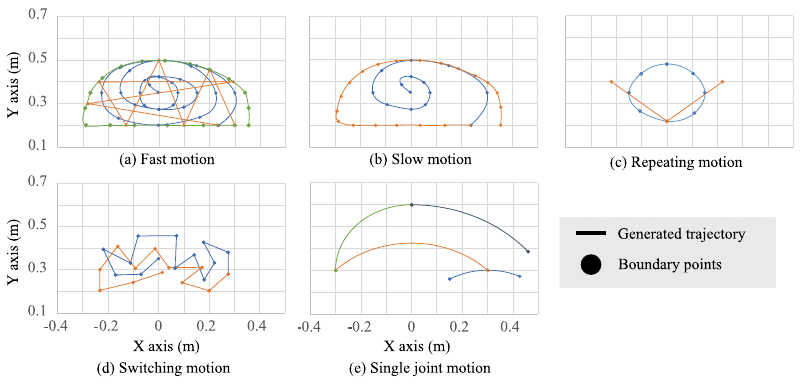}
\caption{The 5 kinds of the designed tasks. The generated trajectory is obtained by minimum jerk model with boundary conditions: x and y positions, velocity, acceleration, and time. Further, These tasks include different motions, which are illustrated as distinct colored trajectories.}
\label{fig:tasks}
\end{figure*}

Subjects conducted 5 kinds of tasks that took approximately 60 s to run, included rest and task parts with 1:1.(Fig. \ref{fig:tasks}). These movements were made by the minimum jerk mode \cite{minimum-jerk-model} with the border conditions. During the experiment, the length of the forearm and upper arm were varied because there is an accuracy error in the sensorless motion capture system. Therefore, we calculate the target marker positions from prepared target angles and the current arm lengths as \eqref{eq:joint22D}.

\begin{equation}
\begin{cases}
x_t^{\mathrm{tgt}} = L_t^{\mathrm{sld}}\cos\bigl(\theta_t^{\mathrm{sld}}\bigr) + L_t^{\mathrm{elb}}\cos\bigl(\theta_t^{\mathrm{sld}} + \theta_t^{\mathrm{elb}}\bigr), \\[2mm]
y_t^{\mathrm{tgt}} = L_t^{\mathrm{sld}}\sin\bigl(\theta_t^{\mathrm{sld}}\bigr) + L_t^{\mathrm{elb}}\sin\bigl(\theta_t^{\mathrm{sld}} + \theta_t^{\mathrm{elb}}\bigr)
\end{cases}
\label{eq:joint22D}
\end{equation}
Where, $x_t^\text{tgt}$ and $y_t^\text{tgt}$ are wrist positions at the $t$ time, $L_t^\text{sld}$ and $L_t^\text{sld}$ are the length of the forearm and upper arm at the $t$ time, and $\theta_t^\text{sld}$ and $\theta_t^\text{elb}$ are the prepared target shoulder and elbow joint angles at the $t$ time. We set the based target points hand-made and extended them by minimum jerk model (frame rate: 120 Hz). We got target joint angles assuming the length of the forearm and upper arm are 30 cm with \eqref{eq:task2joint}.
\begin{equation}
\begin{gathered}
\left\{
\begin{array}{l}
\theta^{\text{sld}} = \operatorname{atan2d}(a,\, b) - \operatorname{atan2d}\Bigl(\sqrt{a^2 + b^2 - c^2},\, c\Bigr)\\[2mm]
\theta^{\text{elb}} = \operatorname{atan2d}\Bigl(\sqrt{a^2 + b^2 - c^2},\, c\Bigr)\\[2mm]
\quad + \operatorname{atan2d}\Bigl(\sqrt{a^2 + b^2 - d^2},\, d\Bigr)
\end{array}
\right.\\[2mm]
\begin{array}{c}
a = y^{wst} - y^{sld}\\[2mm]
b = x^{wst} - x^{sld}\\[2mm]
c = \frac{a^2 + b^2 + {L^\text{sld}}^2 - {L^\text{elb}}^2}{2\, L^\text{sld}}\\[2mm]
d = \frac{a^2 + b^2 - {L^\text{sld}}^2 + {L^\text{elb}}^2}{2\, L^\text{elb}}
\end{array}
\end{gathered}
\label{eq:task2joint}
\end{equation}

We describe how to generate the 5 kinds of the tasks in Fig. \ref{fig:tasks}. These tasks include different motions, which are illustrated as distinct colored trajectories in Fig. \ref{fig:tasks}. For all segments of the motions, the start and end boundary conditions are defined such that the velocity and acceleration of the x and y axes are 0 $\mathrm{rad.}/\mathrm{s}$ and 0 $\mathrm{rad.}/\mathrm{s}^2$, respectively. In particular, for the straight trajectories shown in Fig. \ref{fig:tasks}A, C, and D, the boundary conditions are also set so that the velocity and acceleration in the x and y axes are 0 $\mathrm{rad.}/\mathrm{s}$ and 0 $\mathrm{rad.}/\mathrm{s}^2$. Further, in Fig. \ref{fig:tasks}E, the curved trajectories are got based on the elbow or shoulder joint angle rather than the wrist position.

\section{Motion Labels Processing}
\label{sec: app_motion_labels_processing}

\begin{figure*}[htbp]
\centering
\includegraphics[width=\textwidth]{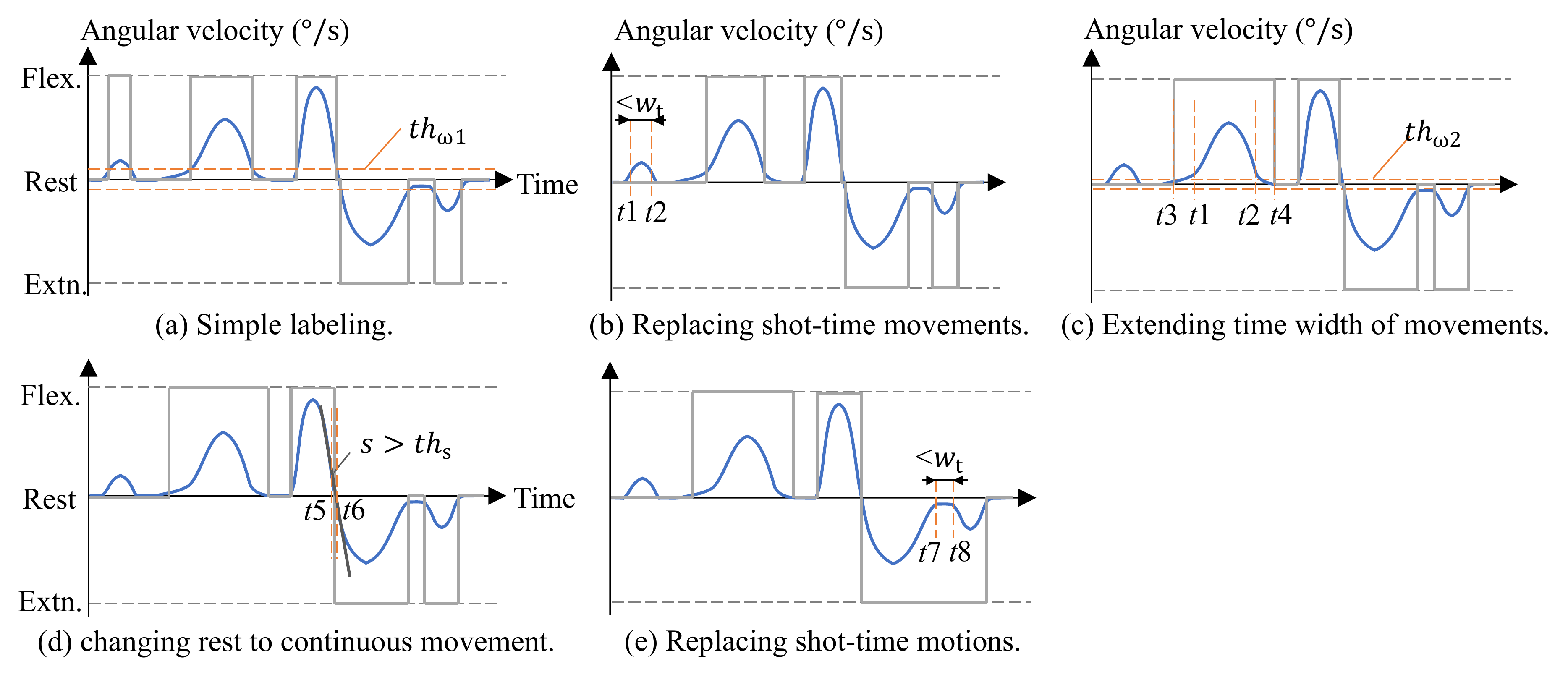}
\caption{The motion labels processing. (a)-(e) indicate processing names. the gray lines mean motion labels. The blue lines mean angular velocity.}
\label{fig:motion_labels_processing}
\end{figure*}

We describe how to get motion labels from elbow joint angular velocity. First, we get the elbow joint angular velocity with \eqref{eq:task2joint} and he difference method. The motion labels are got by 5 steps (Fig. \ref{fig:motion_labels_processing}). $1^\text{st}$ step is motion labeling based on the elbow joint angular velocity by \eqref{eq:motion_labeling}.
\begin{equation}
l_t =
\begin{cases}
\text{flexion}   & (\omega_t > th_{\upomega1}) \\[1ex]
\text{extension} & (\omega_t < -th_{\upomega1}) \\[1ex]
\text{rest}      & (\text{otherwise})
\end{cases}
\label{eq:motion_labeling}
\end{equation}
Here, $l_t$ is a motion label at $t$ time, $\omega_t$ is an elbow joint angular velocity at $t$ time, $th_{\upomega1}$ is a threshold for elbow joint angular velocity at $t$ time. We set $th_{\upomega1}$ is 3.0 $\text{rad./s}$.

$2^\text{nd}$ step replaces the movement labels to rest if the time width of the movement is under $w_t \text{ms}$ by \eqref{eq:motion_padding_rest}.
\begin{equation}
l_{t1 \sim t2} = \text{rest} \quad (t2 - t1 > w_t)
\label{eq:motion_padding_rest}
\end{equation}
Where, $t1$ is the start time for flexion or extension, $t2$ is the end time for flexion or extension. $t1$ and $t2$ are defined by the elbow joint angular velocity is over $th_{\upomega1}$. We set $w_\text{t}$ is $200 \text{ms}$.

$3^\text{rd}$ step extends the movement time width on both ends between $t1$ and $t2$ in order to detect the onset of a movement at an early stage and predict the moment when it fully concludes by \eqref{eq:extending_movement_time}.
\begin{equation}
\begin{aligned}
l_{t3 \sim t1} &= l_{t1} \\
l_{t2 \sim t4} &= l_{t2}
\end{aligned}
\label{eq:extending_movement_time}
\end{equation}
Where, $t3$ is a time that is $|\omega_t|>th_\upomega$ and close to $t1$. $t4$ is a time that is $|\omega_t|>th_\upomega$ and close to $t2$.

$4^\text{th}$ step change the rest to continuous movement in order to prevent the instant at which the type of movement changes from being misclassified as rest by \eqref{eq:contenuelizing_movement}.
\begin{equation}
l_{t5 \sim t6} =
\begin{cases}
\text{flexion}  & (\omega_t \ge th_{\upomega2} \text{ and } s \ge \text{th}_\mathrm{s}) \\[1ex]
\text{extension} & (\omega_t < -th_{\upomega2} \text{ and } s \ge \text{th}_\mathrm{s}) \\[1ex]
\text{unchange}  & (\text{otherwise})
\end{cases}
\label{eq:contenuelizing_movement}
\end{equation}
Where $\omega_t$ is an elbow joint angular velocity at $t$ time, $s$ is a sloop between $t5$ and $t6$, $\text{th}_\mathrm{s}$ is a threshold for the sloop, $t5$ and $t6$ are start and end time for rest. We set $th_{\upomega2}$ is 1.0 $\text{rad./s}$ and $\text{th}_\mathrm{s}$ is $5 \,\mathrm{rad.}/\mathrm{s}^2$ and got $s$ by minimum square method.

The final step replaces the motion labels to the former or post motion label if the time width of the motion label is under $w_t$ ms in order to exclude motions that are shorter than the predefined duration by \eqref{eq:paddig_motions}.
\begin{equation}
l_{t7 \sim t8} =
\begin{cases}
l_{t7-1} & (t7-1 \ge 0) \\
l_{t8+1} & (\text{otherwise})
\end{cases}
\label{eq:paddig_motions}
\end{equation}
Where $t7$ and $t8$ are both ends between one of the motions.

\FloatBarrier
{
\footnotesize
\fontfamily{ptm}\selectfont
\bibliographystyle{IEEEtran}
\bibliography{reference}
}

\end{document}